\documentclass[12pt]{article}

\begin{document}

\newcommand{\be}{\begin{equation}}
\newcommand{\ee}{\end{equation}}
\newcommand{\bea}{\begin{eqnarray}}
\newcommand{\eea}{\end{eqnarray}}
\newcommand{\bi}{\bibitem}
\renewcommand{\r}{({\bf r})}
\renewcommand{\t}{(t)}
\newcommand{\rt}{({\bf r},t)}
\newcommand{\rrpttp}{({\bf r},{\bf r'},t,t')}
\newcommand{\rptp}{({\bf r'},t')}
\newcommand{\ua}{\uparrow}
\newcommand{\da}{\downarrow}

\title{Exploring dynamical magnetism with time-dependent density-functional 
theory: from spin fluctuations to Gilbert damping}

\date{\today}

\author{Klaus Capelle\\
Instituto de Qu\'\i mica de S\~ao Carlos\\
Departamento de Qu\'\i mica e F\'\i sica Molecular\\
Universidade de S\~ao Paulo, Caixa Postal 780\\ 
13560-970 S\~ao Carlos, SP, Brazil\\ \\
Balazs L. Gyorffy\\
H.H. Wills Physics Laboratory\\
University of Bristol, Tyndall Av.\\
Bristol BS8 1TL, U.K.}

\maketitle


\begin{abstract}
We use time-dependent spin-density-functional theory to study dynamical
magnetic phenomena. First, we recall that the local-spin-density approximation
(LSDA) fails to account correctly for magnetic fluctuations in the 
paramagnetic state of iron and other itinerant ferromagnets.
Next, we construct a gradient-dependent density functional that 
does not suffer from this problem of the LSDA. This functional is then used 
to derive, for the first time, the phenomenological Gilbert equation 
of micromagnetics directly from time-dependent density-functional theory.
Limitations and extensions of Gilbert damping are discussed on this basis,
and some comparisons with phenomenological theories and experiments are
made. 
\end{abstract}

The collective behaviour of spins of mobile, correlated electrons that
gives rise to metallic magnetism is still only imperfectly understood.
The problems arising in micromagnetics \cite{micromag} or
spintronics \cite{spintronics}, to mention but two subjects of current
technological interest that involve itinerant magnetism, are therfore 
often dealt with within the framework of
the frankly phenomenological Landau-Lifshitz-Gilbert equations, or
semiphenomenological models whose foundation in material-specific
microscopic theory remains incomplete. On the other hand, first-principles
spin-density-functional theory (SDFT), which gives a more or less adequate
account of the ground state, fails even to address directly the issue of
time dependence of the magnetization. Clearly, under these circumstances it
is worthwhile to inquire what physical insight may be gained by
approaching the problem from the novel point of view afforded by
time-dependent SDFT (TD-SDFT) \cite{tddft,spindynlett}. 
Our aim here is to explore the potential benefits and difficulties of such
enterprise.

In the present paper we thus
(i) explain why the popular local-spin-density approximation (LSDA), on its 
own, cannot account for the magnetic fluctuations at and above the Curie
temperature, not even when combined with TD-SDFT;
(ii) construct a new gradient-dependent density-functional that does not
suffer from this problem of the LSDA; and 
(iii) use this functional to give a microscopic derivation of the 
phenomenological Gilbert equation of micromagnetics. This equation has
recently been invoked for the phenomenological interpretation of 
experiments on magnetic nanolayers \cite{gilbertprl}, but has 
never before been derived microscopically from density-functional theory.

Let us first consider fluctuations in the paramagnetic phase. Among the
low-lying excitations of a ferromagnet above the Curie temperature
are spin fluctuations. Temporarily, in the neighbourhood of an
atom, these fluctuations give rise to magnetic moments that average to
zero on a sufficiently long time scale. A convenient way to describe this 
is by introducing the average magnetization
\be
{\bf \bar{m}}_\tau\r = \frac{1}{\tau} \int _0^\tau dt\, {\bf m}\rt,
\label{average}
\ee
where ${\bf m}\rt$ is the space and time-dependent magnetization density
obtained, in principle, from solving the time-dependent Schr\"odinger
equation for the many-electron problem, and $\tau$ is the averaging time.

Consider now the behaviour of this average magnetization as a function of 
$\tau$. Right after a magnetic fluctuation has given rise to a magnetic
moment on a given site, the average ${\bf \bar{m}}_\tau\r$ will be different 
from zero. Moreover, this moment certainly persists until $\tau$ is of the 
order of the hopping time to the nearest neighbour, $T_{h}$, which can be
estimated by the inverse band width, $\hbar/W$. 
On the other hand, for times much longer than the one characteristic for 
thermal fluctuations of the magnetic moment, $T_{fluc}$,
the average over all configurations of ${\bf m}\rt$ is zero.
These two regimes, one with short-term local moments the other with
no magnetic moments, exist in any ferromagnet above the Curie temperature. 
Depending on the values of band width and Curie temperature there may also 
be a third regime, characterized by averaging times satisfying
$ \hbar/W << \tau << T_{fluc}$. In this case electrons hop from site to site,
but maintain sufficiently strong
correlations to prevent immediate destruction of the local magnetic
moments by thermal fluctuations. This latter regime has been addressed 
by the disordered-local moment picture \cite{dlm} 
in which the electronic hopping is treated with DFT, 
and the much slower behaviour of the local moments is dealt with by
using statistical mechanics.

We now ask whether the above, widely accepted physical picture can be recovered 
within TD-SDFT.
On a formal level there is no reason to expect otherwise, since TD-SDFT is an 
exact transcription of the time-dependent many-body problem, subject only
to quite weak v-representability conditions \cite{robert}, but the situation
is different when one considers the approximations for the density-functional
that are necessary to perform an actual calculation.
The simplest time-dependent density functional is the adiabatic 
LSDA, or ALSDA \cite{tddft}. This functional
is a straightforward generalization
of the conventional LSDA of ground-state DFT, from which it can be
obtained by simply substituting the time-dependent densities $n\rt$
and ${\bf m}\rt$ for the ground-state densities $n\r$ and  ${\bf m}\r$.
 
The first objective of this paper is to point out that a treatment of
magnetic fluctuations within the ALSDA can never correctly account for 
the observed magnetic fluctuations. To understand why, we only have to
consider the TD-SDFT equation of motion for the spin degrees of freedom.
Neglecting spin currents, this equation takes the form \cite{spindynlett}
\be
\frac{\partial {\bf m}\rt}{\partial t}
=-\gamma{\bf m}\rt  \times
\left[ {\bf B}_{ext}\rt + {\bf B}_{xc}\rt\right],
\label{tdsdftlifshitz}
\ee
where  ${\bf B}_{ext}$ is the externally applied magnetic field, $\gamma$ the
gyromagnetic ratio, and
${\bf B}_{xc}\rt$ is the exchange-correlation (xc) magnetic field of TD-SDFT.
Within the ALSDA this field is calculated as
\be
{\bf B}^{ALSDA}_{xc}\rt = \left.
- \frac{\delta E^{LSDA}[n\r,{\bf m}\r]}{\delta {\bf m}\r}
\right|_{n\r\to n\rt, {\bf m}\r \to {\bf m}\rt},
\label{alda}
\ee
where $E^{LSDA}$ is the LSDA functional of static SDFT. Within the LSDA 
${\bf B}_{xc}$ is locally parallel to ${\bf m}$ and hence drops out of 
equation (\ref{tdsdftlifshitz}). In the absence of an externally applied 
magnetic field this equation then predicts a time-independent magnetization 
density ${\bf m}\rt =const$. Within the LSDA the average magnetization
${\bf \bar{m}}_\tau\r$ is thus independent of the averaging
time, and the LSDA is seen to be unable to account for the three different
regimes described above. 
{\it The LSDA thus does not capture the dynamics of magnetic fluctuations.}

This does of course not imply that previous LSDA-based calculations of spin 
fluctuations \cite{dlm,spindyncalc}, which have had considerable empirical 
success, are wrong. Such calcuations typically avoid the problem by 
bringing in concepts from outside of that framework, such as
external constraining fields that are not recalculated selfconsistently, 
adiabaticity assumptions for the spin dynamics, fitting to model Hamiltonians,
identification of Kohn-Sham energies with excitation energies and the
Kohn-Sham susceptibility with the many-body one, etc.
However, it would clearly be desirable to have an approach that is exclusively
based on TD-SDFT, without the need for additional approximations and 
assumptions. This observation provides the motivation for the development, 
in the present paper, of a prototype density functional that 
is designed to avoid the problems of LSDA.

To this end, note first that the problem is an intrinsic defficiency of the 
adiabatic LSDA, which persists even if the functional is formulated in a 
noncollinear way, i.e., in terms of the full magnetization vector ${\bf m}\rt$ 
instead of the more common variables $n_\ua\rt$ and $n_\da\rt$.
In Ref.~\cite{spindynlett} we found that
the simplest way to graft a nontrivial spin dynamics onto the
functional is to include gradient terms. However, not any gradients will
do the job. First of all, the gradients must involve all components of the
magnetization vector. This discards all standard GGA-type functionals,
which are explicitly formulated for the z-component only and can therefore
not account for directional fluctuations of the magnetic moments.
Second, the functional must satisfy the zero-torque theorem 
(ZTT), which states that the net torque exerted by ${\bf B}_{xc}$ on 
the system as a whole must vanish \cite{spindynlett}:
\be
\int d^3r\, {\bf m}\rt  \times {\bf B}_{xc}\rt \equiv 0.
\label{ztt}
\ee

Guided by these considerations we now construct a simple model for a 
${\bf B}_{xc}$ functional that avoids the problems of the LSDA.
Our starting point is the following (still completely general)
representation of ${\bf B}_{xc}$ as
\be
{\bf B}_{xc}\rt = \int d^3r' \int_{-\infty}^t dt' \, 
\hat{K}[n,{\bf m}]\rrpttp {\bf m}\rptp,
\label{kerneldef}
\ee
where $\hat{K}[n,{\bf m}]\rrpttp$ is an, as yet undetermined, tensorial kernel,
characteristic for the spin-spin interactions. To make contact with the
phenomenological theories discussed below we now assume that $\hat{K}$ is a 
short-ranged isotropic function of the form $\hat{K}(|{\bf r}-{\bf r}'|,t-t')$.
This isotropy assumption is reasonable if ${\bf B}_{xc}$ and ${\bf m}$,
and therefore also $\hat{K}$, are interpreted
as averages over intraatomic distances, as is the case in the
phenomenological theories \cite{footnote}. Due to the short-rangedness of 
$\hat{K}$ we can then expand the vector ${\bf m}\rptp$ under the integral in
Eq.~(\ref{kerneldef}) about the point ${\bf r}$ and the time $t$,
\bea
{\bf m}\rptp  =
{\bf m}\rt +
\nonumber \\
({\bf r'}-{\bf r})\nabla\otimes {\bf m}\rt +
\frac{1}{2}[({\bf r'}-{\bf r})\cdot\nabla]^2{\bf m}\rt + 
(t'-t)\dot{\bf m}\rt + \ldots,
\label{mexpand}
\eea
where we have kept first-order terms in the temporal variation,
$\dot{\bf m}\rt = \partial{\bf m}\rt/\partial t$, and second-order
terms in the spatial one. As will become apparent later, it is these orders 
that are required to make contact with phenomenological theories.
Based on experience with similar expansions in static SDFT, nonlinear 
terms in ${\bf m}$ (which would correspond to cubic or higher-order terms
in $E_{xc}$) are not included in this expansion. ${\bf B}_{xc}$ then becomes
\be
{\bf B}_{xc}\rt = \hat{K}_0\t {\bf m}\rt
+ \hat{K}_2\t \nabla^2 {\bf m}\rt
+ \hat{\Lambda}\t \dot{\bf m}\rt,
\label{bxcexpand}
\ee
with
\bea
\hat{K}_0\t = 4\pi \int_{-\infty}^t dt'  \int_0^\infty dx\,x^2 \hat{K}(x,t-t')
\label{k0}
\\
\hat{K}_2\t = 2\pi \int_{-\infty}^t dt'  \int_0^\infty dx\,x^4 \hat{K}(x,t-t')
\label{k2}
\\
\hat{\Lambda}\t = 4\pi \int_{-\infty}^t dt'  \int_0^\infty dx\,x^2 (t'-t) 
\hat{K}(x,t-t'), 
\label{lambda}
\eea
and $x=|{\bf r}-{\bf r}'|$.
Substituting this in the equation of motion (\ref{tdsdftlifshitz}) we obtain
\bea
\dot{\bf m}\rt =
-\gamma{\bf m}\rt\times\left[{\bf B}_{ext}\rt+{\bf B}_{xc}\rt\right]=
\nonumber \\
-\gamma{\bf m}\rt  \times
\left[ {\bf B}_{ext}\rt + 
\hat{K}_0\t {\bf m}\rt +
\hat{K}_2\t \nabla^2 {\bf m}\rt +
\hat{\Lambda}\t \dot{\bf m}\rt
\right].
\label{eqofm}
\eea
Equation (\ref{eqofm}) constitutes a tensorial generalization of the Gilbert 
equation of micromagnetics \cite{micromag,gilbertprl}. 
That equation is usually written as
\be
\dot{\bf m}\rt =
-\gamma{\bf m}\rt  \times {\bf B}_{eff}\rt
+ \frac{\alpha}{M_s}{\bf m}\rt \times \dot{\bf m}\rt,
\label{gilbert}
\ee
where ${\bf B}_{eff}$ is a phenomenological effective field comprising
external and exchange fields, $M_s$ is the saturation magnetization, and
$\alpha$ is the so called {\it Gilbert damping constant}
\cite{micromag,gilbertprl}.
We see that our equation (\ref{eqofm}) reduces to an equation of the form
(\ref{gilbert}) if the kernels we assumed for generality to be tensors
are taken to be scalars, and their intrinsic time-dependence is neglected. 

Interestingly, after the reduction of tensors to scalars the first two terms 
(containing $K_0$ and $K_2$) in ${\bf B}_{xc}$ satisfy the ZTT identically. 
On the other hand, the Gilbert damping term on its own is not
guaranteed to satisfy the ZTT. Since the latter is an exact constraint that
must be obeyed by any effective field to be used in the equation of motion
(\ref{tdsdftlifshitz}), this must be considered a defficiency of the simple
form of the Gilbert damping or, equivalently, the linearization of the
expression (\ref{kerneldef}) for ${\bf B}_{xc}\rt$. Another limitation of
Gilbert theory that becomes apparent from our derivation is that in 
general the coefficients in Eq.~(\ref{eqofm}) are time dependent, while those 
in Eq.~(\ref{gilbert}) are not. This time-dependence can give rise to
additional dynamics and damping that is not described by the Gilbert term. 
Below we will show that there is a simple model within which one can 
rationalize this lack of intrinsic time dependence of the coefficients.

Carrying on, momentarily, with time-independent scalars $K_0$ and $K_2$
instead of 
time-dependent tensors, we can write the ${\bf B}_{xc}$ functional as
\be
{\bf B}_{xc}\rt = K_0 {\bf m}\rt + K_2 \nabla^2 {\bf m}\rt
+ \Lambda \dot{\bf m}\rt,
\label{bxcfunc2}
\ee
where $K_0$, $K_2$, and $\Lambda$ are numbers that characterize the 
spin-spin interaction kernel. The correspondence between the 
phenomenological and the microscopic equations, found above, shows one way 
in which these numbers can be obtained, since the parameters entering the 
Gilbert equation can be extracted from experiment or simulations 
\cite{micromag}:
The first term in Eq.~(\ref{bxcfunc2}) is of the form obtained in the
LSDA, i.e., parallel to ${\bf m}$. (This observation implies that one
can, in principle, determine $K_0$ from the LSDA.) Interestingly, the 
derivative
terms in the expression (\ref{bxcfunc2}) imply that the resulting
${\bf B}_{xc}$ is not parallel to ${\bf m}$ even when the coefficients
are reduced from tensors to scalars. As a consequence they give, unlike the
LSDA, rise to nontrivial spin dynamics and spin fluctuations when substituted 
back into the equation of motion. 
To show why we expect the resulting spin dynamics to be essentially
correct, in spite of the approximations made in the argument so far,
let us, temporarily, neglect the damping term, and write 
${\bf B}_{eff}\rt = {\bf B}_{ext}\rt  + K_2 \nabla^2 {\bf m}\rt$
(the term containing $K_0$ does not contribute to the equations of motion
in the scalar approximation).
This is precisely of the form of the effective field entering the
phenomenological Landau-Lifshitz equation \cite{lali}, which is known
to correctly account for the observed spin dynamics in ferromagnets.
In this context $K_2$ is usually replaced by the spin wave stiffness
$D$, which is related to our $K_2$ by $D=\gamma \hbar M_s K_2$ and
can be obtained from experiment or independent calculations
(see, e.g., Ref.~\cite{schwitalla} and references therein).

One way to completely fix the parameters in our functional is thus
to determine $K_0$ from the LSDA, $K_2$ from the spin stiffness, 
and $\Lambda$ from the Gilbert damping constant. 
Clearly, this empiricism diminishes the predictive power of the functional, 
and we will propose below a slightly less empirical strategy in its place.
As it stands, the main virtue of the construction leading to 
Eqs.~(\ref{eqofm}) and (\ref{bxcfunc2}) is rather in its conceptual
consequences:
(i) it gives some degree of microscopic justification to the phenomenological
approaches, and (ii) it shows what the term missing in the LSDA treatment
of spin dynamics and spin fluctuations is, and how a simple prototype 
${\bf B}_{xc}$ functional that has this term might look like.

The simple functional (\ref{bxcfunc2}) is already sufficient to recover 
the two basic regimes
for spin fluctuations above the Curie temperature, discussed in the 
introduction. To show this we substitute Eq.~(\ref{bxcfunc2}) into
(\ref{eqofm}) and linearize by setting ${\bf m}\rt = {\bf M}_0 + {\bf n}\rt$,
where ${\bf n}$ describes small fluctuations about the equilibrium
magnetization ${\bf M}_0$. The resulting set of coupled differential equations 
for ${\bf n}$ is readily identified as describing damped oscillations with 
frequency
\be
\Omega(q) = \sqrt{\omega(q)^2-{k(q)^2 \over 4}} + i {k(q) \over 2},
\ee
where
$\hbar\omega(q)=\hbar\gamma(B+q^2 M_s K_2)/(1+\alpha^2)
=(g \mu_0 B+ D q^2)/(1+\alpha^2)$, $B=|{\bf B}_{ext}|$, and 
$k(q)=2\alpha\omega(q)/(1+\alpha^2)$.
Below the Curie temperature ${\bf M}_0$ is the dominating contribution to
${\bf m}\rt$ and to the average ${\bf \bar{m}}_\tau$ of Eq.~(\ref{average}).
Above this temperature ${\bf M}_0=0$, and the $\tau$-averaged magnetic moment
is entirely due to the damped spin fluctuations described by ${\bf n}\rt$.
As a consequence of the damping, the average magnetic moment is zero for
sufficiently large $\tau$. On a time scale shorter than that set by the
damping constant $\Lambda$, however, the contribution of 
${\bf n}\rt$ to ${\bf \bar{m}}_\tau$ does not vanish, and one obtains a 
transient magnetic moment. This fundamental separation of time scales is
not obtained from the LDA, within which $\Lambda = 0$ and there is no damping.
Furthermore, in the absence of the external field ${\bf B}_{ext}$ the spin
dynamics continues to be driven by the gradient term $\propto K_2$, 
whereas in the LDA $K_2=0$ and there is no intrinsic spin dynamics at all
($\Omega^{LDA}\equiv 0$).

We now proceed to give a somewhat more microscopic characterization of the 
coefficients $K_0$, $K_2$ and $\Lambda$. To this end we recall that according 
to Eqs.~(\ref{k0}) to (\ref{lambda}) these coefficients are all defined
as certain integrals over the spin-spin interaction kernel 
$K(|{\bf r}-{\bf r}'|,t-t')$.
A simple, but not unrealistic, model for the space and time dependence
of this kernel is 
\be
K(x,t)=K e^{-x/\lambda} e^{-t/T},
\label{model}
\ee
which amounts to assuming that spin-spin interactions decay exponentially
both in space and time.
The amplitude $K$ in Eq.~(\ref{model}) is a measure for the strength of the
spin-spin interaction, while $\lambda$ measures its spatial range and $T$
its temporal range, i.e., its memory.
With this model for the spin-spin kernel the integrals (\ref{k0}) to 
(\ref{lambda}) can be evaluated analytically, and one finds 
$K_0 = 8\pi KT\lambda^3$,
$K_2 = 48 \pi KT\lambda^5 = 6 \lambda^2 K_0$, and
$\Lambda = -8\pi K \lambda^3 T^2 = -T K_0$.
These relations have several interesting consequences.
First of all, they replace the coefficients $K_0$, $K_2$ and $\Lambda$, 
which had been introduced above in a purely mathematical way, by the
physically meaningful quantities $K$, $\lambda$ and $T$.
Second, although the kernel $K(x,t)$ in Eq.~(\ref{model}) has an explicit
time dependence, the coefficients calculated with it depend only on the
memory time $T$, and not on $t$ itself. Within this model we can thus
justify the replacement of time-dependent coefficients by static ones,
made by the phenomenological theories [cf. our discussion of the limitations
of conventional Gilbert damping, below Eq.~(\ref{gilbert})].
On the other hand, recent experiments report enhanced Gilbert damping in
$Fe$-$Au$ nanolayers \cite{gilbertprl}. In layered systems the
isotropic exponential model (\ref{model}) for $K$ is, of course, unrealistic, 
and its main conclusion, the time-independence of $K_0$, $K_2$ and $\Lambda$, 
does not hold. The microscopic equation (\ref{eqofm}) then retains a 
time-dependence in its coefficients, and attempts to recast it in the form
of the traditional Gilbert equation (with a time independent $\Lambda \propto 
\alpha$) must result in the appearance of additional damping terms. 
Interestingly, additional damping has indeed been found necessary in various
proposals \cite{gilbertprl} for explaining the experimental results.

Returning now to the model (\ref{model}), we note that the expressions 
obtained from it for $K_0$, $K_2$, amd $\Lambda$ can be used, together with 
the above connection of these coefficients with the phenomenological parameters 
in the Landau-Lifshitz-Gilbert equations, to deduce relations
between the latter and the microscopic parameters $K$, $\lambda$ and $T$.
An interesting example is
\be
\lambda = \sqrt{\frac{T D}{6 \hbar \alpha}}
= \sqrt{\frac{D}{6\gamma \hbar M_s K_0}},
\ee
which shows how the microscopic quantities $\lambda$ and $T$ (characterizing 
the spatial range and memory of the spin-spin interaction 
kernel) are related to the macroscopic parameters $\alpha$ (Gilbert damping), 
$M_s$ (saturation magnetization) and $D$ (spin stiffness).
To test the consistency of these relations we now connect them to
another semiphenomenological approach, namely Stoner theory.
Recall that in that theory there are no gradient-dependent terms,
since it can be interpreted as a linearized LSDA. If we use it
to evaluate $\lambda$ we thus expect that the gradient-dependent terms in
our functional disappear. This is indeed the case: In Stoner theory the
proportionality factor between the effective magnetic field and the local
magnetization is just $1/2\mu_0^2$ times the Stoner parameter $I$. Hence 
$K_0 \approx I/(2\mu_0^2)$. Plugging this into the preceeding equation and
using for $I$, $D$, and $M_s$ the experimental values for iron, we find that
$ \lambda = 0.18\times 10^{-10}m$. This value, which is much shorter than
the lattice constant in iron ($2.7 \times 10^{-10}m$),
shows that Stoner theory is indeed a local theory, in which spin-spin
interactions decay very rapidly. It also shows, through the relation
$K_2=6 \lambda^2 K_0$, that the gradient-dependent terms in our functional
are strongly suppressed when one uses Stoner theory to evaluate them, as
expected.

Finally, we point out that the time average of our expression for 
${\bf B}_{xc}$, Eq.~(\ref{bxcfunc2}), can also be interpreted as a 
magnetic equation of state \cite{spinfluc}, and used to extract the 
equilibrium magnetization. Naturally, in the paramagnetic state the 
time-dependence of ${\bf B}_{xc}\rt$ is due to fluctuations only, and 
the long-time average of ${\bf B}_{xc}\rt$ is zero. In the ferromagnetic 
state, on the other hand, this average remains finite. Upon including 
a cubic term in ${\bf m}$ in the expansion and calculating the average 
according to Eq.~(\ref{average}), we indeed recover the usual equations 
of spin-fluctuation theory \cite{spinfluc}, determining the equilibrium 
magnetization and the Curie temperature. Here, however, they are obtained 
by invoking the ergodic theorem, characterizing equilibrium via temporal 
averaging instead of by thermal averages. 

The main results of this work are as follows: We have shown that the LSDA 
fails to account for spin fluctuations near the Curie temperature, even 
when it is made time-dependent (ALSDA) and fully noncollinear. We next 
identified certain gradient-dependent terms as the key missing ingredient
in the ALSDA, and constructed a very simple
prototype functional that contains these gradients. Although simple,
this functional gives rise to a still quite general equation of motion
for the spin degrees of freedom, which can be identified as a
microscopic version of the equations of motion of Landau-Lifshitz and
of Gilbert.
This identification provides microscopic support for these phenomenological
theories and opens up a way for the empirical determination of the parameters
in the functional. However, it also brings to light some shortcomings of
the phenomenological approaches, which may be relevant for recent experiments.
In summary, we have presented evidence that time-dependent SDFT may be a
novel and useful alternative to constrained SDFT as a first-principles 
description of itinerant magnetism in metals.

{\bf Acknowledgments} KC acknowledges useful discussions with J.~Quintanilla
and financial support from FAPESP. BLG is pleased to acknowledge a 
R\"ontgen Professorship at the University of W\"urzburg, during the 
tenure of which some of the above work was done.

\end{document}